\documentclass[conference]{IEEEtran}
\IEEEoverridecommandlockouts
\usepackage{amsmath,amssymb,amsfonts}
\usepackage{graphicx}
\usepackage{textcomp}
\usepackage{booktabs} 
\usepackage{makecell}
\usepackage{multirow}
\usepackage{adjustbox}
\usepackage{hyperref}
\usepackage{url}
\usepackage{nicefrac}
\usepackage{microtype}
\usepackage{xcolor}
\usepackage{makecell}
\usepackage{algorithm}
\usepackage{algpseudocode}
\usepackage{tikz}
\usepackage{stfloats}
\usetikzlibrary{positioning,arrows.meta,shapes.geometric,calc}

\usepackage{orcidlink}
\usepackage{comment}
\usepackage{braket}
\def\BibTeX{{\rm B\kern-.05em{\sc i\kern-.025em b}\kern-.08em
    T\kern-.1667em\lower.7ex\hbox{E}\kern-.125emX}}
\usepackage{cite}
\usepackage{pifont}
\usepackage{amsmath,amssymb,amsfonts}
\usepackage{graphicx}
\usepackage{textcomp}
\usepackage{amsmath}
\usepackage{booktabs} 
\usepackage{color}
\usepackage{multirow}
\usepackage{colortbl, xcolor}

\usepackage{hyperref}       
\usepackage{url}            
\usepackage{booktabs}       
\usepackage{amsfonts}       
\usepackage{nicefrac}       
\usepackage{microtype}      
\usepackage{xcolor}         
\usepackage{algorithm}
\usepackage{algpseudocode}
\usepackage[skip=5pt]{caption}

\usepackage{amsmath}
\usepackage{tikz}

\usepackage{orcidlink}
\usepackage{comment}
\usepackage{amssymb}
\usepackage{braket}
\usepackage{xcolor}
\usepackage{setspace}

\usepackage{enumitem}
\usepackage{braket}
\usepackage{xcolor}
\def\BibTeX{{\rm B\kern-.05em{\sc i\kern-.025em b}\kern-.08em
    T\kern-.1667em\lower.7ex\hbox{E}\kern-.125emX}}
\begin{document}
\bstctlcite{bstctl:etal, bstctl:nodash, bstctl:simpurl}

\title{Robustness Evaluation of Hybrid Quantum \\Neural Networks under Noise Models\\ via System-Level Error Mitigation}

\author{
\IEEEauthorblockN{Jesse Roberta Mingue Njiki\textsuperscript{1}, Nouhaila Innan\textsuperscript{2,3}, Alberto Marchisio\textsuperscript{2,3,4}$^*$\thanks{$^*$This work has been carried out while at NYUAD.},
Muhammad Kashif\textsuperscript{2,3},\\ Jean-Michel Dricot\textsuperscript{1}, and Muhammad Shafique\textsuperscript{2,3}}
\vspace{2mm}
\IEEEauthorblockA{\textsuperscript{1}\normalsize Université Libre de Bruxelles, Av. Roosevelt 50, 1050 Bruxelles, Belgium\\
\textsuperscript{2}\normalsize eBRAIN Lab, Division of Engineering, New York University Abu Dhabi (NYUAD), Abu Dhabi, UAE\\
\textsuperscript{3}\normalsize Center for Quantum and Topological Systems (CQTS), NYUAD Research Institute, NYUAD, Abu Dhabi, UAE\\
\textsuperscript{4}\normalsize SDU Microelectronics, Institute of Mechanical and Electrical Engineering, University of Southern Denmark, Odense, Denmark\\
jesseming2017@gmail.com, jean-michel.dricot@ulb.be, chisio@sdu.dk,\\
\{nouhaila.innan, muhammadkashif, muhammad.shafique\}@nyu.edu}}


\maketitle

\begin{abstract}
Quantum Neural Networks (QNNs) represent a promising direction within Quantum Machine Learning (QML), yet their realization on noisy intermediate-scale quantum (NISQ) devices remains constrained by decoherence, gate imperfections, crosstalk, and readout errors. This study provides a systematic evaluation of noise effects and mitigation strategies in hybrid quantum neural networks (HQNNs). Zero-Noise Extrapolation (ZNE), Digital Dynamical Decoupling (DDD), and Layerwise Richardson Extrapolation (LRE) are incorporated into end-to-end QNN training pipelines developed with PennyLane, simulated using Qiskit Aer noise models, and integrated with the Mitiq framework, while Probabilistic Error Cancellation (PEC) is evaluated separately under depolarizing noise due to its computational cost. Experiments conducted on the Iris dataset with five representative noise channels show that the impact of noise and the effect of mitigation are strongly dependent on the noise model and its strength. The model maintains comparatively strong performance under phase-flip and phase-damping noise, while substantial degradation is observed under high depolarizing and amplitude-damping noise. Across the evaluated mitigation methods, the observed benefits remain limited and noise-dependent: ZNE, DDD, and LRE generally follow the same degradation trends as the unmitigated baseline, while PEC shows limited gains only in the low-noise depolarizing regime. These findings highlight the need for context-specific mitigation strategies to improve the robustness of QNNs in practical NISQ settings.
\end{abstract}


\section{Introduction}

\begin{figure}[t!]
    \centering
    \includegraphics[width=1\linewidth]{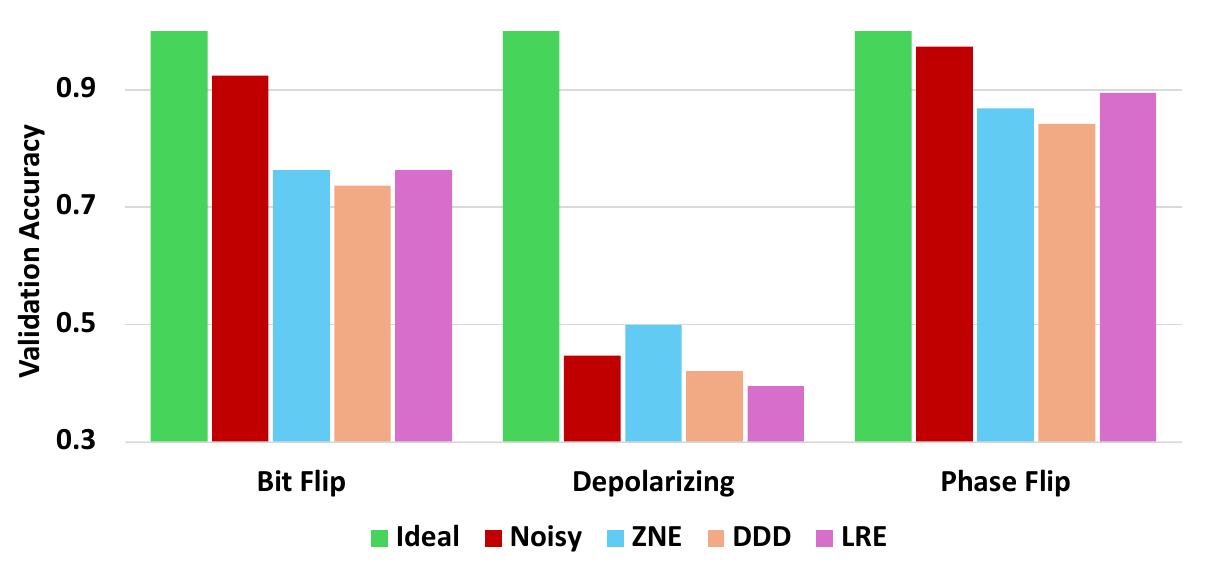}
    \caption{Motivational analysis illustrating the behavior of the QNN model under various noise models (Bit Flip, Depolarizing, and Phase Flip) and mitigation techniques (ZNE, DDD, LRE). The results reveal that mitigation does not uniformly enhance performance across all scenarios, emphasizing the need to investigate which techniques are most effective for specific noise types and model configurations.}
    \label{motiva}
\end{figure}

Quantum computing has emerged as a disruptive paradigm, exploiting superposition, entanglement, and interference to address problems that are intractable for classical systems. Within this field, Quantum Machine Learning (QML) has gained momentum, with Quantum Neural Networks (QNNs) using Parameterized Quantum Circuits (PQCs) that mimic the layered structure of classical neural networks to offer the potential for enhanced data representation and learning capabilities~\cite{surveyqml2023,ciliberto2018quantum,qnnreview2023,Kashif:demonstrating,chang2025primer, kashif2025computational,innan2025next}. 
Despite this promise, the practical realization of QNNs is constrained by the limitations of Noisy Intermediate-Scale Quantum (NISQ) devices, which suffer from decoherence, gate infidelity, crosstalk, and readout noise that significantly degrade learning stability, convergence, and accuracy~\cite{latif2023survey,benchmarking2023, ahmed2025quantum, kashif2024investigating,ahmed2025noisyhqnn}. 
As our preliminary experiments confirm (see Fig.~\ref{motiva}), even shallow QNN architectures experience rapid performance deterioration under depolarizing or phase-flip noise, raising fundamental concerns about their reliability in high-assurance domains such as IoT security, finance, and anomaly detection. Interestingly, mitigation techniques do not always yield improvement and can even exacerbate degradation, highlighting the complex interplay between noise type, circuit depth, and mitigation strategy. This observation forms a key motivation for this study.

Existing studies on Quantum Error Mitigation (QEM) have concentrated largely on quantum algorithms such as the Variational Quantum Eigensolver (VQE) or the Quantum Approximate Optimization Algorithm (QAOA)~\cite{cai2022quantumerrormitigation,zhang2023benchmarking}. The unique learning dynamics of QNNs, characterized by gradient vanishing, circuit expressibility, and stochastic noise sensitivity, demand dedicated analyses that account for both algorithmic and hardware constraints. \textit{This gap underscores the need to understand how different mitigation strategies behave under various noise channels, and how they affect the overall training process in practical QML scenarios.}
To address this research gap, we present a comprehensive experimental study of QEM techniques integrated within hybrid QNN workflows. 

\textbf{The main contributions of this work are as follows:}
 \begin{itemize}
 \item \textit{A unified analysis framework that incorporates four leading QEM techniques}, namely Zero Noise Extrapolation (ZNE), Probabilistic Error Cancellation (PEC), Digital Dynamical Decoupling (DDD), and Layerwise Richardson Extrapolation (LRE), \textit{to simulate and train QNNs under noisy conditions}.
\item \textit{Evaluation of 128 experimental configurations, repeated over 384 total runs, across five representative noise channels} (depolarizing, amplitude damping, phase damping, bit flip, and phase flip) using the Iris dataset, providing \textit{an extensive benchmarking of mitigation strategies in full QNN learning pipelines}.
 \item \textit{A comparative evaluation of the techniques in terms of accuracy and convergence stability}, revealing distinct performance patterns across noise models and highlighting trade-offs between accuracy gains and resource overhead.
 \end{itemize}

\textbf{Summary of Key Results:} Our systematic analysis shows that QNN accuracy degradation depends on both the type and strength of the applied noise, with additional variation across mitigation settings. The model maintains comparatively strong performance under phase-flip and phase-damping noise, while substantial degradation is observed under high depolarizing and amplitude-damping noise. Across the evaluated mitigation methods, the observed benefits remain limited and noise-dependent: ZNE, DDD, and LRE generally follow the same degradation trends as the unmitigated baseline, while PEC, evaluated only under depolarizing noise, shows limited gains in the low-noise regime.

\vspace{4pt}

\textbf{Paper Organization:} Sec.~\ref{sec:related} reviews existing research on QNNs and quantum error mitigation. Sec.~\ref{sec:methodology} details the proposed experimental framework and simulation setup. Sec.~\ref{sec:results} presents and analyzes the empirical results, highlighting observed trends across noise and mitigation models. Finally, Sec.~\ref{sec:conclusion} summarizes the key insights and outlines future directions for adaptive QEM strategies in practical quantum learning systems.

\section{Background and Related Work \label{sec:related}}

\begin{figure}[t!]
    \centering
    \includegraphics[width=1\linewidth]{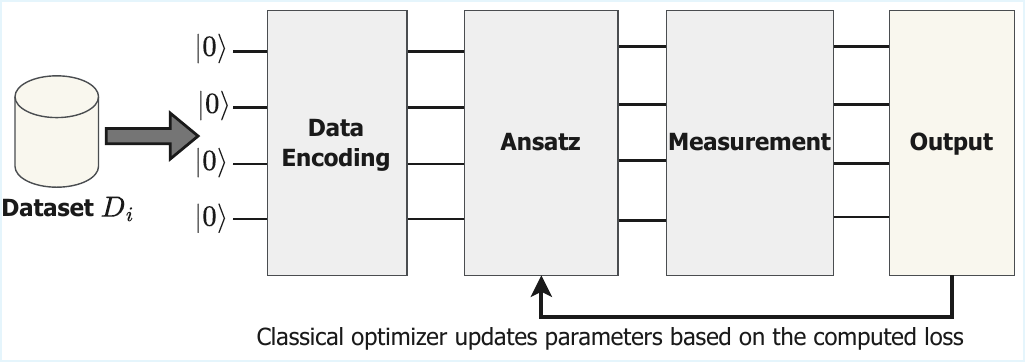}
    \caption{General architecture of QNNs, illustrating data encoding, parameterized quantum circuit (ansatz), and measurement process. The classical optimizer updates parameters based on the computed loss, forming a hybrid training loop.}
    \label{qnn}
\end{figure}

\begin{table*}[hbtp]
\centering
\caption{Common single-qubit noise channels.}
\label{tab:noise-models}
\renewcommand{\arraystretch}{1.2}
\begin{adjustbox}{max width=\textwidth}
\begin{tabular}{l l l}
\hline
\textbf{Noise model} & \textbf{Equation} & \textbf{Description} \\
\hline

Depolarizing &
\(\rho' = (1 - p)\rho + \frac{p}{3}\left(X\rho X + Y\rho Y + Z\rho Z\right)\) &
\makecell[l]{A noise process in which the qubit state is left unchanged with probability \(1-p\), and with total\\ probability
\(p\), one of the Pauli operators is applied uniformly at random. This channel models\\ isotropic
randomization of the qubit state caused by imperfect gate operations and calibration drift.} \\
\hline

Amplitude Damping &
\makecell[l]{\(\rho' = E_0 \rho E_0^\dagger + E_1 \rho E_1^\dagger\)\\
\(E_0=\begin{bmatrix}1&0\\0&\sqrt{1-p}\end{bmatrix},\;
E_1=\begin{bmatrix}0&\sqrt{p}\\0&0\end{bmatrix}\)} &
\makecell[l]{Energy relaxation, where an excited state \(\ket{1}\) decays to the ground state \(\ket{0}\) due to spontaneous\\
emission. This model captures \(T_1\) relaxation effects that limit qubit coherence time in\\ superconducting
and trapped-ion technologies.} \\
\hline

Phase Damping &
\makecell[l]{\(\rho' = E_0 \rho E_0^\dagger + E_1 \rho E_1^\dagger\)\\
\(E_0=\begin{bmatrix}1&0\\0&\sqrt{1-p}\end{bmatrix},\;
E_1=\begin{bmatrix}0&0\\0&\sqrt{p}\end{bmatrix}\)} &
\makecell[l]{Loss of quantum coherence without affecting energy populations. It attenuates the off-diagonal\\ terms of the
density matrix while preserving the diagonal elements. This type of noise is\\ commonly associated with \(T_2\)
processes and phase uncertainty in superconducting-qubit systems.} \\
\hline

Bit Flip &
\(\rho' = (1 - p)\rho + p\,X\rho X\) &
\makecell[l]{Inversion of a qubit's logical value, flipping \(\ket{0}\) to \(\ket{1}\) or vice versa, with probability \(p\).\\ This error type is analogous
to classical bit corruption and often originates from control-pulse\\imperfections or environmental crosstalk.} \\
\hline

Phase Flip &
\(\rho' = (1 - p)\rho + p\,Z\rho Z\) &
\makecell[l]{Randomly applies a \(Z\) operation with probability \(p\), introducing a phase inversion between\\ computational
basis states. It disrupts interference in quantum superpositions while leaving\\ measurement probabilities in the
computational basis unchanged.} \\
\hline

\end{tabular}
\end{adjustbox}
\end{table*}

\subsection{Quantum Neural Networks}
QNNs combine the principles of quantum computing with the layered architecture of classical neural networks, employing PQCs to learn data representations and generate predictions~\cite{ciliberto2018quantum,qnnreview2023,kashif2021design,el2026comparative}. As shown in Fig.~\ref{qnn}, a QNN typically consists of three main stages: quantum data encoding, parameterized unitary evolution (ansatz), and quantum measurement. Classical data are first encoded into quantum states, transformed by trainable quantum gates, and then measured to produce outcomes used to compute a loss function. A classical optimizer updates the circuit parameters based on the computed loss, forming a hybrid quantum–classical training loop.
In hybrid architectures, additional classical layers are often integrated before or after the quantum circuit to enhance flexibility and performance. Classical preprocessing layers, such as convolutional or linear layers, can reduce input dimensionality and extract relevant features before quantum encoding, thereby lowering circuit depth and mitigating hardware noise. Conversely, post-quantum layers, such as a linear transformation followed by a softmax activation, are frequently employed for classification or regression, translating quantum measurement outputs into normalized probability distributions for comparison with target labels.
 
QNNs have demonstrated the potential to represent complex functions that may be intractable for classical networks, enabling promising applications across chemistry, finance, and cybersecurity~\cite{abbas2020power,biamonte2018qml,pathak2024resource,innan2025qnn,dave2025sentiqnf,11228312,innan2024financial,innan2024quantum,choudhary2025hqnn,innan2025optimizing,dutta2025quiet,innan2025lep}. Despite this potential, QNN training remains challenging due to gradient vanishing phenomena (barren plateaus), optimization instability, and sensitivity to noise in near-term quantum devices~\cite{jeswal2018qnnreview,disparatenoise2024,kashif2024resqnets,kashif2023impact,kashif2025deep,atallah2025investigating}. To realistically assess QNN performance and reliability under such hardware limitations, noise effects are often simulated using quantum noise models that replicate common error sources observed in NISQ processors\cite{kashif2024hqnet,kashif2024nrqnn}. These models provide a framework for studying the impact of decoherence, gate infidelity, and measurement errors on the learning dynamics of QNNs.

\subsection{Quantum Noise Models}
\begin{table*}[b]
\centering
\caption{Comparison of related QEM benchmarking studies and this work.}
\label{tab:sota_comparison}
\resizebox{\linewidth}{!}{
\begin{tabular}{@{}lllll@{}}
\toprule
\textbf{Reference} & \textbf{Target System} & \textbf{Techniques Tested} & \textbf{Noise Models} & \textbf{QNN Focus} \\ \midrule
Cai et al. (2023)~\cite{cai2022quantumerrormitigation} & VQE/QAOA & ZNE, PEC & Depolarizing, Amplitude Damping & \ding{55} 
 \\
Weaving et al. (2023)~\cite{zhang2023benchmarking} & VQE & ZNE, PEC, CDR & Depolarizing, Pauli & \ding{55} 
 \\
Russo et al. (2024)~\cite{russo2024layerwise} & Variational Circuits & LRE & Depolarizing, Damping & \ding{55} 
 \\
Krebsbach et al. (2022)~\cite{krebsbach2022optimized} & General Variational & ZNE (optimized) & Depolarizing & \ding{55} 
 \\
\textbf{This Work} & \textbf{Hybrid QNN (PennyLane–Qiskit)} & \textbf{ZNE, PEC, DDD, LRE} & \textbf{5 Noise Types} & \textbf{\ding{51}} \\ \bottomrule
\end{tabular}
}
\end{table*}

Quantum noise arises from the inevitable interaction between a quantum system and its surrounding environment, leading to decoherence and the loss of quantum information. In NISQ devices, such effects are unavoidable and must be modeled to realistically capture the non-ideal behavior of quantum circuits. Quantum noise is typically described using quantum channels, represented by completely positive trace-preserving (CPTP) maps acting on the system's density matrix~$\rho$. These channels can be expressed using the Kraus operator formalism as $\rho' = \sum_i E_i \rho E_i^\dagger,$ where $\sum_i E_i^\dagger E_i = I,$ with $\rho$ denoting the density matrix of the quantum system before the noise interaction, $\rho'$ denoting the resulting (noisy) density matrix after the application of the channel, and ${E_i}$ denoting the Kraus operators that represent the effect of the environment on the system. Each operator $E_i$ corresponds to a possible noise event, and the completeness relation $\sum_i E_i^\dagger E_i = I$ ensures that the transformation is CPTP, guaranteeing the physical validity of the quantum operation.

Different physical noise sources are characterized by distinct sets of Kraus operators and corresponding parameters, most commonly a noise probability~$p$ representing the likelihood of an error event. The most representative quantum noise channels simulated in NISQ environments are discussed in Table~\ref{tab:noise-models}.

\subsection{Quantum Error Mitigation}

To alleviate the detrimental effects of quantum noise, various error suppression and mitigation strategies have been developed. While full quantum error correction remains beyond the reach of current NISQ hardware, QEM techniques provide practical alternatives for improving the reliability of quantum computations without requiring additional qubits or complex syndrome measurements. These techniques estimate or suppress noise effects at the software level, striking a balance between hardware feasibility and computational accuracy~\cite{cai2022quantumerrormitigation, zneunified2023}.

Among the most studied QEM strategies are ZNE, PEC, DDD, and LRE. Each of these methods aims to approximate the noise-free expectation value of observables but differs in principle and resource cost.

\begin{enumerate}
    \item \textbf{ZNE} is one of the most hardware-efficient approaches for NISQ devices. Instead of reducing noise directly, ZNE estimates the noise-free output of a quantum circuit by artificially amplifying the noise and extrapolating the results to the zero-noise limit using polynomial or Richardson extrapolation models~\cite{giurgica2021zne, krebsbach2022optimized}. Since it requires no auxiliary qubits and integrates easily with hybrid learning frameworks, ZNE has become widely adopted in variational algorithms and QNNs.
    \item \textbf{PEC} provides a theoretically exact method for noise mitigation by statistically inverting the noise channel. It represents noisy operations as linear combinations of ideal operations weighted by quasiprobabilities~\cite{vandenberg2022pec_sparse}. While PEC yields unbiased estimates of noise-free observables, it incurs an exponentially increasing sampling overhead as circuit complexity or noise strength grows, making it computationally demanding for deep or large circuits.
    \item \textbf{DDD} mitigates decoherence by inserting carefully designed sequences of identity-equivalent gate operations (decoupling pulses) between quantum gates. These sequences symmetrize qubit-environment interactions, suppressing low-frequency and correlated noise~\cite{ezzell2022dd_survey, ddd2024}. DDD requires no extra qubits and is particularly effective against dephasing and energy relaxation noise but offers limited benefits for control or crosstalk errors.
    \item \textbf{LRE} extends the ZNE principle by performing extrapolation independently across circuit layers rather than globally~\cite{russo2024layerwise}. This layer-specific mitigation enhances numerical stability in deep variational circuits, such as QNNs or QAOA-like architectures, while reducing the overall number of additional circuit evaluations required.
\end{enumerate}

While all these approaches contribute to improving quantum computation fidelity, they differ in scalability and resource efficiency. In the context of QNNs, where repeated circuit executions are required for gradient-based optimization, techniques such as ZNE and LRE are often preferred due to their relatively low overhead compared to PEC, which remains computationally intensive.

To situate recent advancements within a broader research perspective, Table~\ref{tab:sota_comparison} summarizes representative benchmarking works addressing QEM performance, noise modeling, and circuit-level reliability. Most prior studies have primarily focused on variational algorithms such as VQE and QAOA, evaluating a limited subset of noise channels and excluding QNN-specific analyses. In contrast, our study extends this scope by integrating multiple mitigation strategies, including ZNE, PEC, DDD, and LRE, under diverse noise environments within a unified hybrid QNN setting. This distinction underscores the need for systematic benchmarking of QEM techniques in learning-oriented quantum models, emphasizing both robustness and trainability under realistic NISQ conditions.

\section{Methodology \label{sec:methodology}}
\begin{figure*}[ht]
 \centering
 \includegraphics[width=1.0\linewidth]{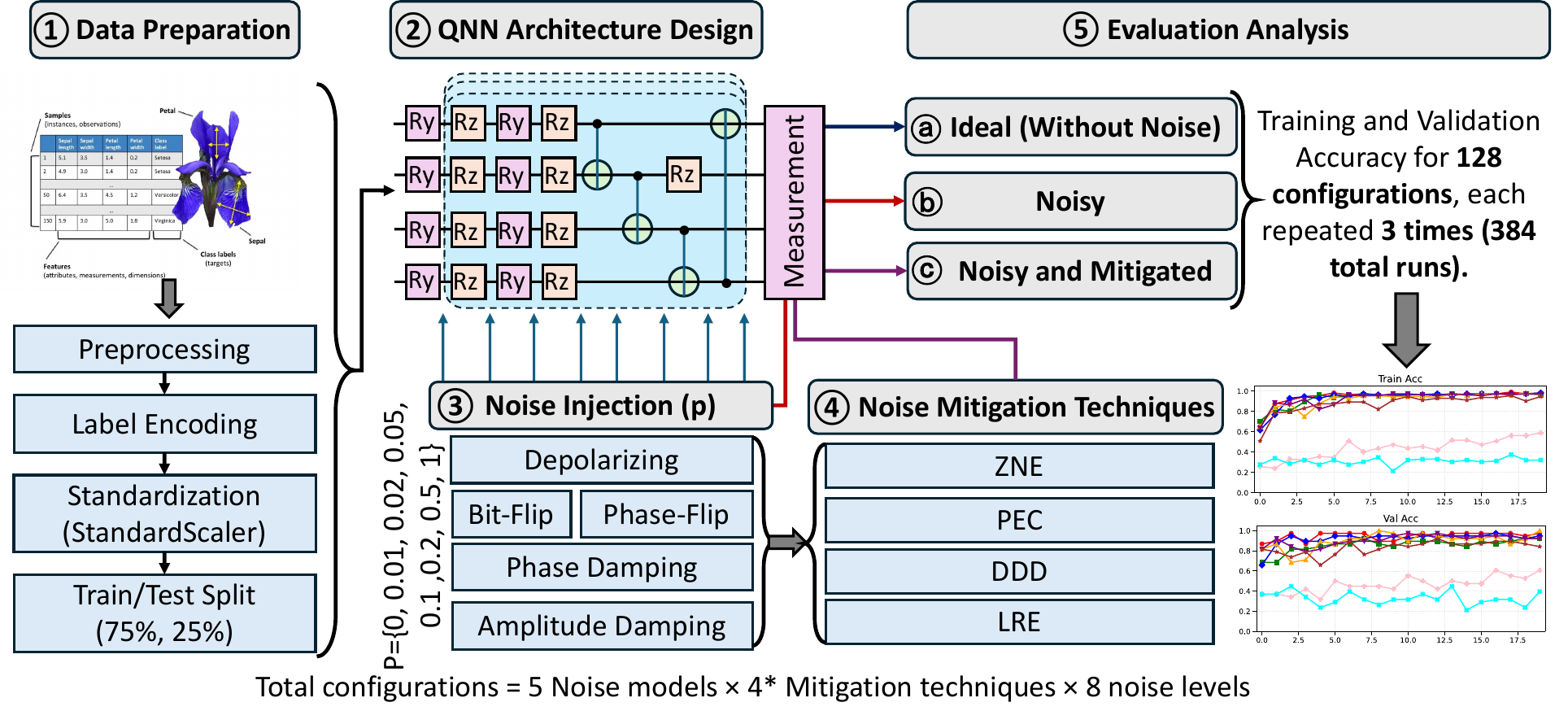}
\caption{Overview of the proposed methodology for benchmarking QNN robustness under noise. The framework considers five noise channels and four mitigation strategies across multiple noise strengths. ZNE, DDD, and LRE are evaluated over all five noise models, whereas PEC is evaluated only under depolarizing noise due to its computational cost. This results in 128 total configurations across the reported experiments.}
 \label{fig:methodology}
 \end{figure*}
We design a modular and reproducible experimental framework to evaluate the robustness of Quantum Neural Networks (QNNs) under realistic noise conditions.
As illustrated in Fig.~\ref{fig:methodology}, the workflow integrates five main stages: (1) Data preparation, (2) QNN architecture design, (3) Noise injection, (4) Noise mitigation, and (5) Evaluation analysis.
The study is conducted using the Iris dataset \cite{Dua:2019}. All experimental parameters are kept consistent across runs, with only the noise model and mitigation strategy varied. The data are preprocessed through label encoding and standardization (\texttt{StandardScaler}), followed by a 75/25 train-test split to ensure balanced class representation.

\subsection{QNN Architecture Design}

The QNN is implemented in \texttt{PennyLane} using three qubits and four variational layers.
Input features from the Iris dataset are encoded through $R_Y$ rotations via \texttt{AngleEmbedding}, followed by four \texttt{StronglyEntanglingLayers} composed of parameterized single-qubit rotations and controlled-Z (CZ) entangling gates.
Measurements are performed in the $Z$ basis, and the results are passed through a classical linear layer with a softmax activation for classification.
In total, the hybrid model contains 36 trainable quantum parameters, balancing expressivity and circuit depth to remain within the NISQ limit.

\subsection{Quantum Noise Injection}

 We simulate five representative quantum noise channels: depolarizing, amplitude damping, phase damping, bit flip, and phase flip, using the \texttt{Qiskit Aer} simulator.
 Noise probabilities $p \in \{0, 0.01, 0.02, 0.05,$ $0.1, 0.2, 0.5, 1.0\}$ are applied before and after each quantum gate in the transpiled circuit, emulating realistic gate-level errors.
 Each simulation uses 8192 measurement shots to approximate hardware sampling noise, and all runs are averaged over three repetitions to ensure statistical stability.

\subsection{Error Mitigation Techniques}

 To address the performance degradation induced by noise, we integrate four QEM techniques from the \texttt{Mitiq} library: 
ZNE, PEC, DDD, and LRE \cite{mitiqdocs}.
 Each technique is applied within the same QNN training loop to either extrapolate ideal outcomes or suppress decoherence effects.
 All runs use identical seeds, data splits, and hyperparameters to isolate the effect of each mitigation method.

\subsection{Training and Evaluation}

All models are trained for 20 epochs using the Adam optimizer with an initial learning rate of 0.3, which is reduced by half every 5 epochs to ensure stable convergence, with a batch size of 5 samples and 3 repetitions per configuration to obtain averaged and statistically consistent results.
To capture the impact of different noise conditions, training is conducted under three experimental settings: (a) ideal (without noise), (b) noisy (with noise but no mitigation), and (c) noisy and mitigated (with one of the four QEM techniques). 
Performance is evaluated based on training and validation accuracy, cross-entropy loss, and runtime overhead, providing a comprehensive view of both learning behavior and computational cost. Overall, the framework spans 128 configurations, combining five noise models and eight noise levels across ZNE, DDD, and LRE, with PEC evaluated only under depolarizing noise, offering a thorough assessment of QNN robustness and the effectiveness of each mitigation strategy.

\section{Results and Discussion \label{sec:results}}

All experiments are conducted on a workstation equipped with an Intel Core i7 (12-thread) CPU and 16 GB of RAM, running Ubuntu 22.04 LTS with Python 3.9. The software stack integrates \texttt{PennyLane}, \texttt{Qiskit Aer}, \texttt{Mitiq v0.45.1}, \texttt{PyTorch}, and \texttt{scikit-learn}.

\subsection{Baseline Performance under Noise}

\begin{figure}[t!]
 \centering
 \includegraphics[width=1.0\linewidth]{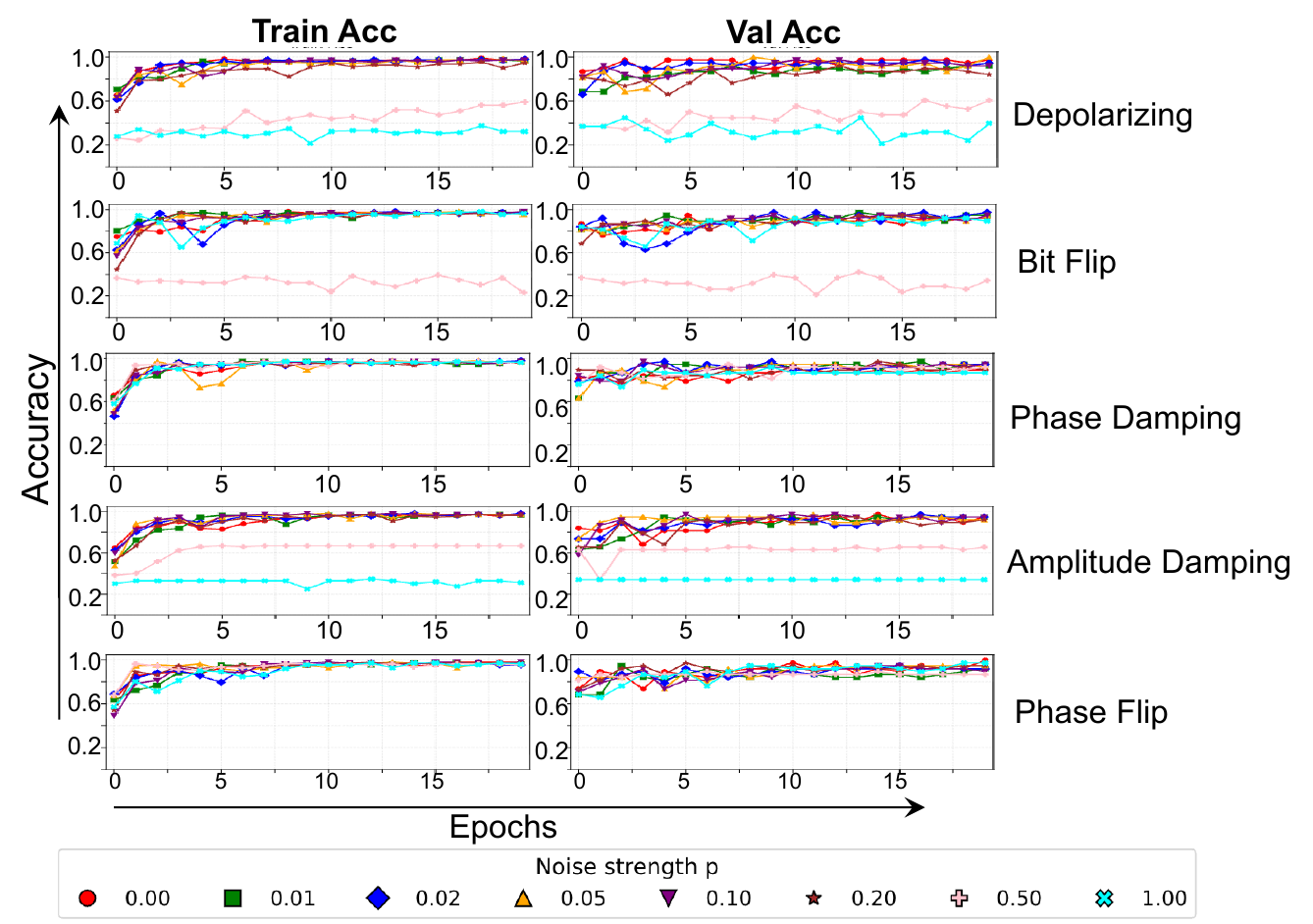}
 \caption{Baseline QNN validation accuracy without mitigation under different noise types.}
 \label{fig:baseline_results}
\end{figure}

Under noiseless conditions, the QNN achieves over 95\% validation accuracy, confirming that the circuit architecture and training configuration are well-suited for the Iris classification task. As shown in Fig.~\ref{fig:baseline_results}, accuracy declines as noise strength increases, with distinct degradation patterns across the different channels. Depolarizing and amplitude-damping noise produce the strongest degradation, with performance dropping more clearly at higher noise levels. Bit-flip noise remains relatively stable at low and moderate noise strengths, but shows a more visible decrease at higher noise levels. In contrast, phase-flip and phase-damping noise are less disruptive, with the model maintaining comparatively strong performance under these channels. A possible explanation is that depolarizing and amplitude-damping noise affect both population and coherence, whereas phase-related errors primarily affect relative phases, which the QNN may partially compensate for through subsequent rotations.

\subsection{Effect of Error Mitigation Techniques}

We next examine how each QEM strategy responds to the applied noise models, focusing on its influence on learning stability and accuracy.

\begin{figure}[h]
    \centering
    \includegraphics[scale=0.6]{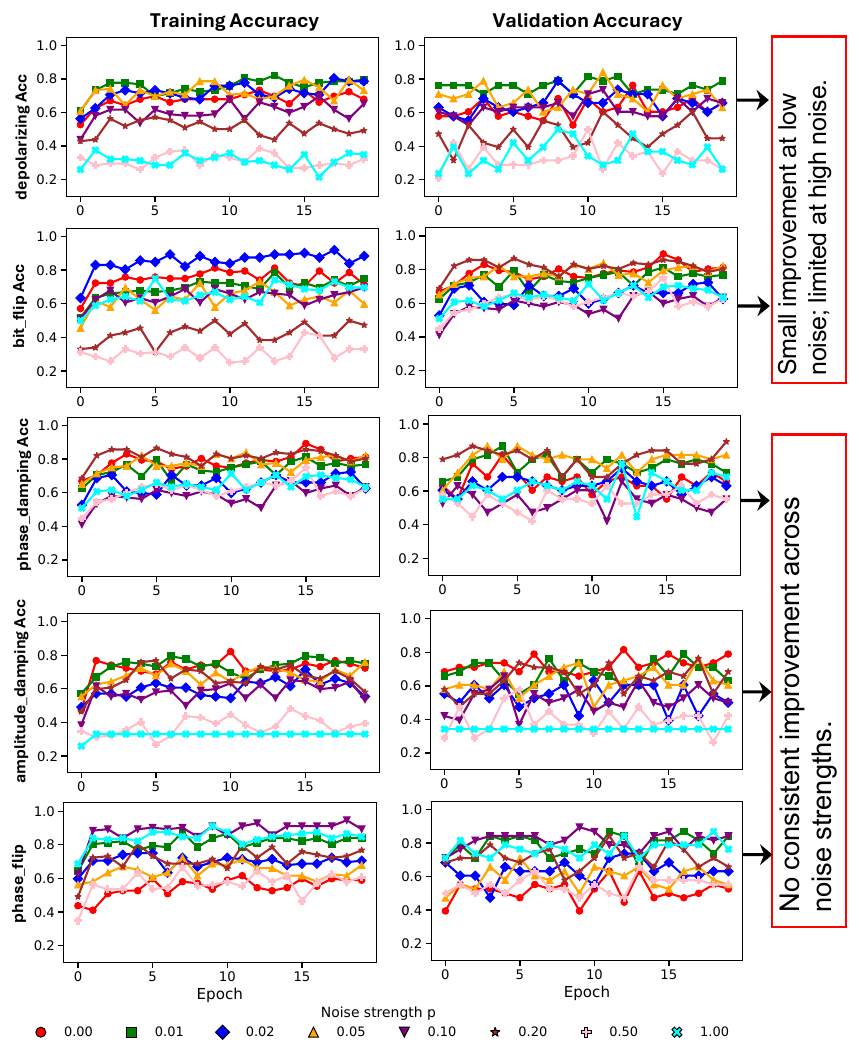}
    \caption{Comparison of training and validation accuracy for ZNE versus the baseline under five noise types.}
    \label{fig:zne_vs_baseline}
\end{figure}
\subsubsection{ZNE}
As shown in Fig.~\ref{fig:zne_vs_baseline}, ZNE exhibits limited and inconsistent benefits across the evaluated noise models. For depolarizing noise, its performance generally follows the same degradation trend as the unmitigated baseline as noise strength increases. For bit-flip noise, ZNE shows small selective improvements in validation accuracy at some noise levels, although these gains are not consistent across training and validation metrics. Under phase-flip, phase-damping, and amplitude-damping noise, ZNE does not provide clear or sustained recovery over the baseline, especially at higher noise levels. These results suggest that, within our experimental setting, ZNE offers only modest and noise-dependent benefits rather than robust mitigation across all regimes.
\subsubsection{PEC}
\begin{figure}[t!]
    \centering
     \includegraphics[width=\linewidth]{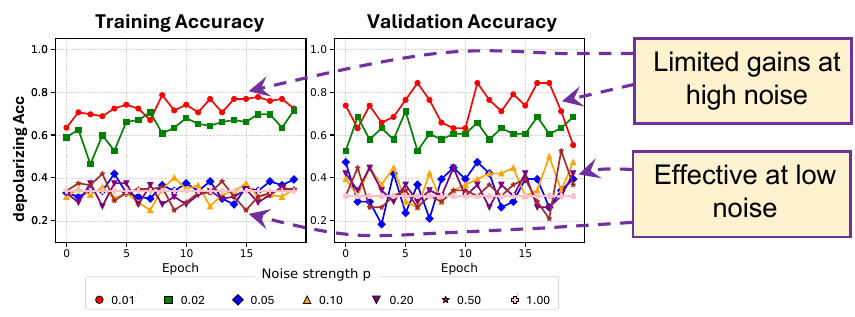}
    \caption{Validation accuracy comparison for PEC and baseline under depolarizing noise.}
    \label{fig:pec_vs_baseline}
\end{figure}

PEC was evaluated only under depolarizing noise, as shown in Fig.~\ref{fig:pec_vs_baseline}. In this setting, PEC follows the same general degradation trend as the unmitigated baseline: performance remains competitive at very low noise levels, but decreases as the noise strength increases. Although PEC is theoretically well suited to Pauli-type noise models such as depolarizing noise, our results show only limited benefit in practice within the present experimental setup. For this reason, and due to its high computational cost in terms of shot budget and runtime, PEC was not extended to the other noise models. Overall, these results suggest that PEC provides only limited benefit in the weak-noise regime for depolarizing errors and was not practical to evaluate more broadly in our simulation pipeline. 
\subsubsection{DDD}
\begin{figure}[t!]
    \centering
    \includegraphics[scale=0.52]{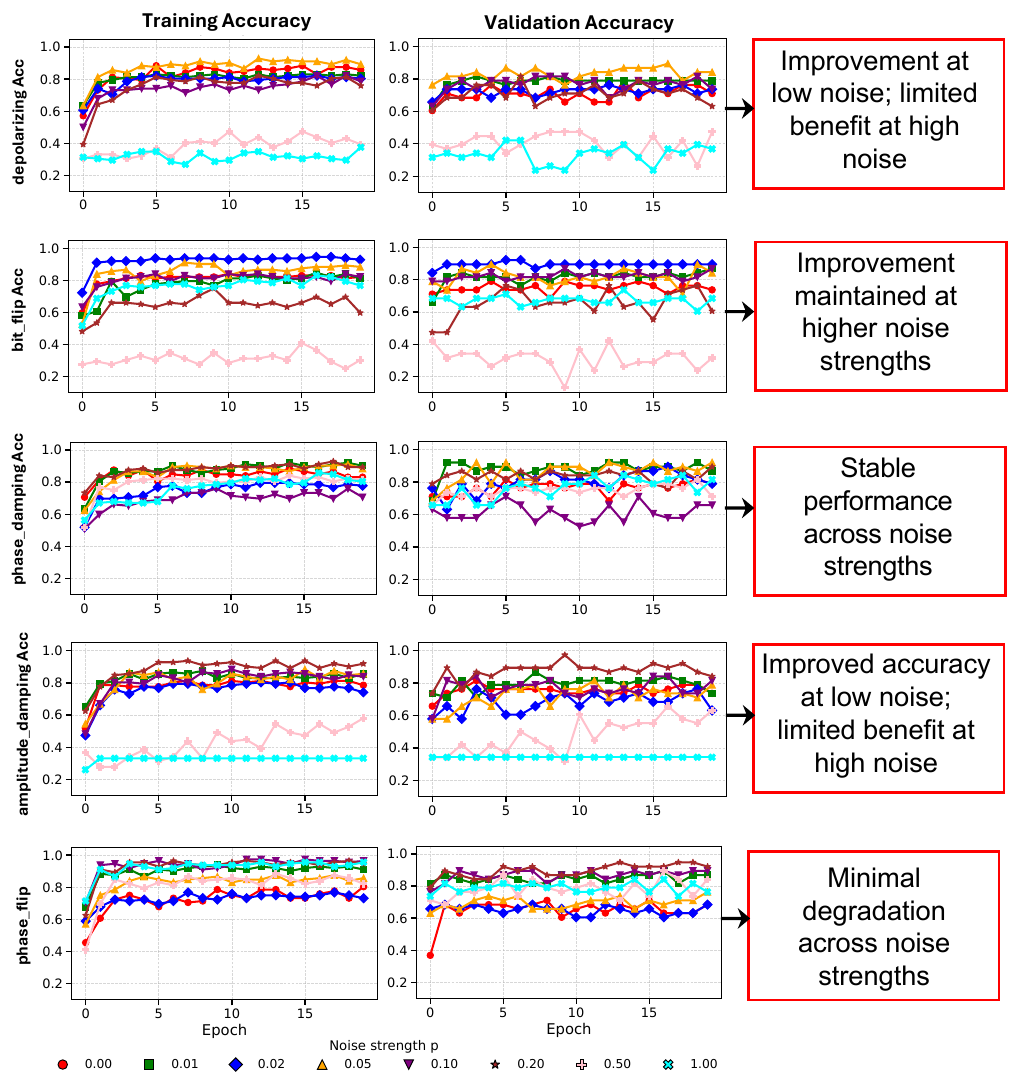}
    \caption{DDD versus baseline accuracy trends across representative noise types.}
    \label{fig:ddd_vs_baseline}
\end{figure}

As shown in Fig.~\ref{fig:ddd_vs_baseline}, DDD provides limited and noise-dependent effects across the evaluated noise models. Under depolarizing and amplitude-damping noise, it does not prevent the clear degradation observed at higher noise strengths. Under bit-flip, phase-flip, and phase-damping noise, DDD maintains relatively good performance at lower noise levels, although this behavior remains close to that of the baseline. These results suggest that DDD may offer modest benefits in selected settings, but it does not consistently restore performance across all noise regimes.

\subsubsection{LRE}
\begin{figure}[t!]
    \centering
    \includegraphics[scale=0.5]{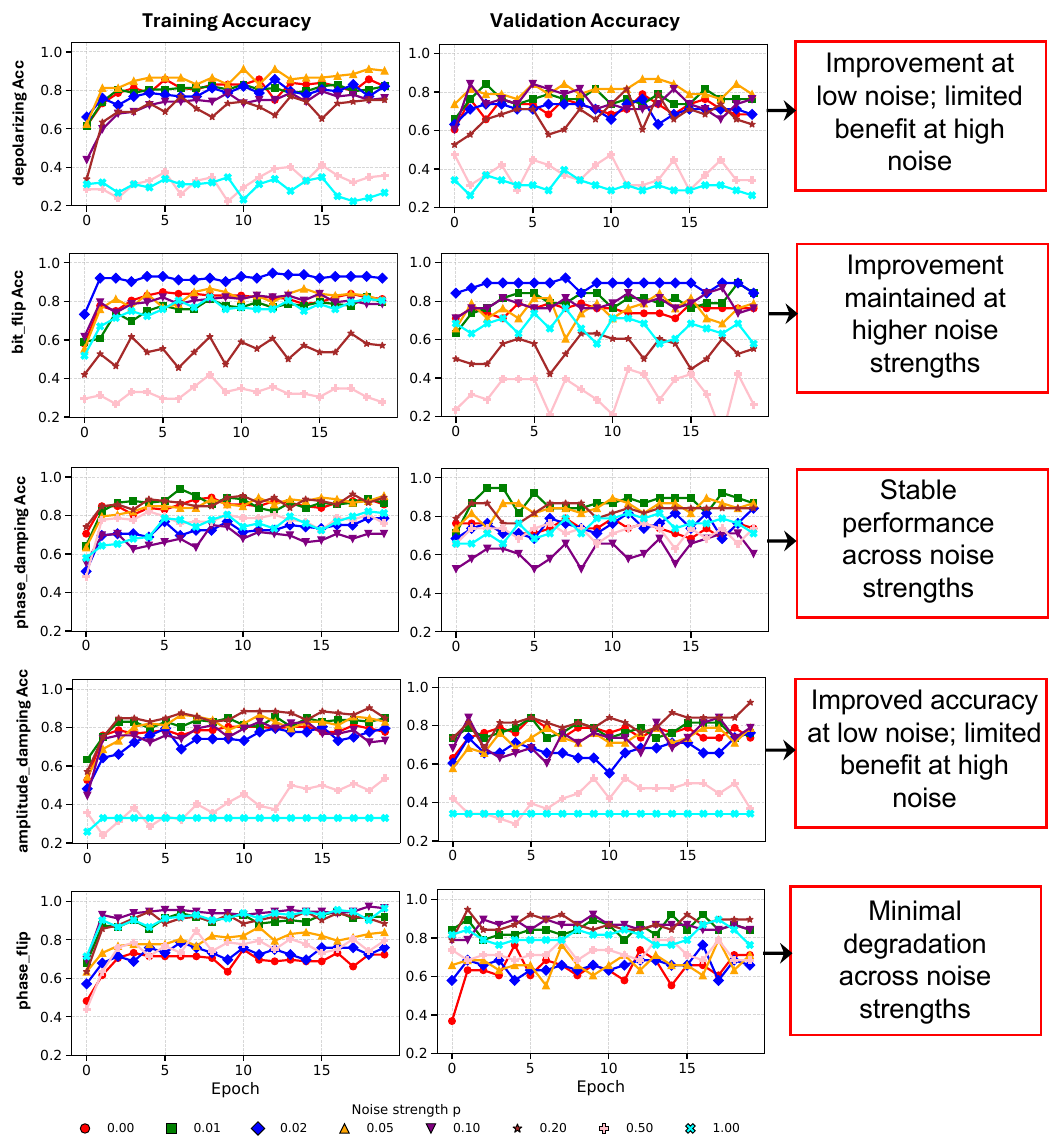}
    \caption{LRE versus baseline under multiple noise channels.}
    \label{fig:lre_vs_baseline}
\end{figure}

As shown in Fig.~\ref{fig:lre_vs_baseline}, LRE exhibits limited and selective effects across the evaluated noise models. For depolarizing and amplitude-damping noise, performance still degrades as noise strength increases, and LRE does not provide clear recovery at higher noise levels. For bit-flip noise, its behavior remains close to that of the baseline, including at higher noise levels where no clear mitigation effect is observed. Under phase-flip and phase-damping noise, the model continues to maintain relatively good performance, but this trend is also already present in the unmitigated setting. Therefore, within our experimental setup, LRE should be interpreted as providing modest and noise-dependent benefits rather than consistent accuracy recovery.
\subsection{Parametric Analysis and Discussion}

The combined results across noise levels, channel types, and mitigation strategies reveal several important behaviors regarding the sensitivity of QNNs in NISQ environments (see Table \ref{tbalenoise}). Rather than showing identical degradation trends across all settings, the QNN exhibits noise-dependent behavior in which both the \emph{type} and the \emph{strength} of the disturbance influence performance. 

In particular, depolarizing and amplitude-damping noise produce the clearest degradation as noise increases, whereas phase-flip and phase-damping noise lead to comparatively smaller performance variations. Bit-flip noise shows an intermediate behavior, with relatively stable performance at low and moderate noise levels and more visible degradation at higher noise strengths. These results suggest that QNN robustness depends not only on the nominal error rate, but also on how the noise process interacts with the circuit structure and the encoded task.
\begin{table}[t!]
\centering
\large
\caption{Accuracy across noise types, noise-strength ranges, and mitigation techniques. Noise strengths are grouped as: Low (0--0.02), Medium (0.05--0.1), and High (0.2--1.0).}
\begin{adjustbox}{max width=\linewidth}
\begin{tabular}{l c c c c c c}
\toprule
\textbf{Noise Type} & \textbf{Range} & \textbf{No Mitigation} & \textbf{DDD} & \textbf{LRE} & \textbf{ZNE} & \textbf{PEC} \\
\midrule
\multirow{3}{*}{\textbf{Depolarizing}} 
& Low    & \textbf{1.0000} & 0.8158 & 0.8420 & 0.8158 & 0.8421 \\
& Medium & \textbf{0.9737} & 0.8947 & 0.8680 & 0.8421 & 0.5000 \\
& High   & \textbf{0.6053} & 0.4737 & 0.4740 & 0.5000 & 0.5263 \\
\midrule
\multirow{3}{*}{\textbf{Bit Flip}}
& Low    & \textbf{0.9737} & 0.9211 & 0.9210 & 0.8947 & --- \\
& Medium & \textbf{0.9474} & 0.8947 & 0.8680 & 0.7368 & --- \\
& High   & \textbf{0.9211} & 0.7368 & 0.7630 & 0.7632 & --- \\
\midrule
\multirow{3}{*}{\textbf{Phase Flip}}
& Low    & \textbf{1.0000} & 0.8947 & 0.9210 & 0.8684 & --- \\
& Medium &\textbf{ 0.9737} & 0.9474 & 0.9210 & 0.8947 & --- \\
& High   & \textbf{0.9737} & 0.8947 & 0.9470 & 0.8684 & --- \\
\midrule
\multirow{3}{*}{\textbf{\makecell{Amplitude\\ Damping}}}
& Low    & \textbf{0.9737} & 0.8684 & 0.8420 & 0.8158 & --- \\
& Medium & \textbf{0.9737} & 0.9737 & 0.9210 & 0.7632 & --- \\
& High   & \textbf{0.6579} & 0.6579 & 0.5260 & 0.5789 & --- \\
\midrule
\multirow{3}{*}{\textbf{\makecell{Phase\\ Damping}}}
& Low    & \textbf{0.9737} & 0.9211 & 0.9470 & 0.8684 & --- \\
& Medium & \textbf{0.9737} & 0.8947 & 0.8680 & 0.8947 & --- \\
& High   & \textbf{0.9474} & 0.8421 & 0.8160 & 0.7632 & --- \\
\bottomrule
\end{tabular}
\end{adjustbox}
\label{tbalenoise}
\end{table}
The evaluated mitigation strategies do not provide uniform recovery across the considered settings. Instead, their effects remain limited and strongly dependent on the underlying noise model and noise strength. ZNE, LRE, and DDD generally follow the same overall degradation trends as the unmitigated baseline, with only small and selective deviations in some cases. PEC, evaluated only under depolarizing noise, also remains close to the baseline trend and shows limited benefit beyond the low-noise regime. As a result, the present experiments do not support a broad claim that any single mitigation method consistently restores QNN performance across all conditions.

The empirical trends across all noise models and mitigation techniques can be summarized as follows:
\begin{itemize}
    \item \textbf{Noise type and noise strength jointly shape performance degradation}, with depolarizing and amplitude-damping noise having the strongest impact in our setting.
    \item \textbf{No single mitigation technique} provides consistent improvement across all evaluated conditions.
    \item \textbf{Mitigation effects remain selective and limited}: small benefits may appear in specific regimes, but none of the tested methods provides systematic recovery under strong noise.
\end{itemize}

The interaction between noise characteristics, circuit structure, and mitigation behavior suggests that designing resilient QNNs requires more than selecting a single mitigation method. Instead, it requires matching circuit design and mitigation choices to the dominant physical error processes of the target hardware. These observations also motivate future work on adaptive or hybrid approaches that treat different noise mechanisms separately rather than assuming a uniform mitigation strategy. Possible directions include noise-aware ansatz design, selective protection of the most sensitive layers, and adaptive switching between mitigation modes during training.
\section{Conclusion \label{sec:conclusion}}

This work presented a system-level evaluation of QEM strategies for HQNNs operating in noisy NISQ settings. The study integrated ZNE, DDD, and LRE within a unified QNN training pipeline built using PennyLane, Qiskit Aer, and Mitiq, while PEC was evaluated separately only under depolarizing noise due to its computational cost. 

Experiments conducted on the Iris dataset under five representative quantum noise models, namely depolarizing, amplitude damping, phase damping, bit flip, and phase flip, revealed channel-dependent degradation patterns. In particular, depolarizing and amplitude-damping noise produced the strongest performance degradation as noise increased, whereas phase-damping and phase-flip noise were less disruptive. 

Across the evaluated mitigation methods, the observed benefits were limited and strongly noise-dependent. ZNE, DDD, and LRE generally followed the same degradation trends as the unmitigated baseline, with only small selective deviations in some settings. PEC, evaluated only for depolarizing noise, showed limited benefit in the low-noise regime while remaining computationally expensive.

The comparative results indicate that QNN robustness cannot be reliably recovered by any single mitigation method in the present setting. Instead, the findings highlight that the effectiveness of mitigation depends on the interaction between the noise model, its strength, and the circuit structure. The proposed evaluation framework therefore provides a useful and reproducible basis for studying the interplay between noise, mitigation, and trainability in hybrid quantum-classical systems.

Future work will extend these analyses to real quantum hardware, deeper circuit topologies, and adaptive mitigation strategies that better account for different physical error mechanisms. Such directions are important for improving the reliability of near-term QNNs in practical QML applications.
\section*{Acknowledgment}
 This work was supported in part by the NYUAD Center for Quantum and Topological Systems (CQTS), funded by Tamkeen under the NYUAD Research Institute grant CG008, and the Center for Cyber Security (CCS), funded by Tamkeen under the NYUAD Research Institute Award G1104.
\bibliographystyle{IEEEtran}

\bibliography{refs}

@IEEEtranBSTCTL{bstctl:nodash, CTLdash_repeated_names = "no", }

@article{ahmed2025quantum,
  title={A comparative analysis and noise robustness evaluation in quantum neural networks},
  author={Ahmed, Tasnim and Kashif, Muhammad and Marchisio, Alberto and Shafique, Muhammad},
  journal={Scientific Reports},
  volume={15},
  number={1},
  pages={33654},
  year={2025},
  publisher={Nature Publishing Group UK London}
}

@inproceedings{kashif2024investigating,
  title={Investigating the effect of noise on the training performance of hybrid quantum neural networks},
  author={Kashif, Muhammad and Sychiuco, Emman and Shafique, Muhammad},
  booktitle={2024 International Joint Conference on Neural Networks (IJCNN)},
  pages={1--10},
  year={2024},
  organization={IEEE}
}

@inproceedings{kashif2025computational,
  title={Computational advantage in hybrid quantum neural networks: Myth or reality?},
  author={Kashif, Muhammad and Marchisio, Alberto and Shafique, Muhammad},
  booktitle={2025 62nd ACM/IEEE Design Automation Conference (DAC)},
  pages={1--7},
  year={2025},
  organization={IEEE}
}

@article{kashif2023impact,
  title={The impact of cost function globality and locality in hybrid quantum neural networks on nisq devices},
  author={Kashif, Muhammad and Al-Kuwari, Saif},
  journal={Machine Learning: Science and Technology},
  volume={4},
  number={1},
  pages={015004},
  year={2023},
  publisher={IOP Publishing}
}

@INPROCEEDINGS{Kashif:demonstrating,
  author={Kashif, Muhammad and Al-Kuwari, Saif},
  booktitle={2022 IEEE International Conference on Rebooting Computing (ICRC)}, 
  title={Demonstrating Quantum Advantage in Hybrid Quantum Neural Networks for Model Capacity}, 
  year={2022},
  volume={},
  number={},
  pages={36-44},
  doi={10.1109/ICRC57508.2022.00011}}

@article{ciliberto2018quantum,
  title={Quantum machine learning: a classical perspective},
  author={Ciliberto, Carlo and Herbster, Mark and Ialongo, Alessandro Davide and Pontil, Massimiliano and Rocchetto, Andrea and Severini, Simone and Wossnig, Leonard},
  journal={Proceedings of the Royal Society A: Mathematical, Physical and Engineering Sciences},
  volume={474},
  number={2209},
  pages={20170551},
  year={2018},
  publisher={The Royal Society Publishing}
}

@article{surveyqml2023,
  title={A survey on quantum machine learning: Current trends, challenges, opportunities, and the road ahead},
  author={Zaman, Kamila and Marchisio, Alberto and Hanif, Muhammad Abdullah and Shafique, Muhammad},
  journal={arXiv preprint arXiv:2310.10315},
  year={2023}
}

@article{chang2025primer,
  title={A primer on quantum machine learning},
  author={Chang, Su Yeon and Cerezo, M},
  journal={arXiv preprint arXiv:2511.15969},
  year={2025}
}

@article{qnnreview2023,
  title={A Review of Quantum Neural Networks: Methods, Models, Dilemmas},
  author={Renxin Zhao and Shi Wang},
  journal={Engineering Reports},
  year={2021},
  eprint={arXiv:2109.01840}
}

@inproceedings{latif2023survey,
  title={A survey and tutorial on security and resilience of quantum computing},
  author={Saki, Abdullah Ash and Alam, Mahabubul and Phalak, Koustubh and Suresh, Aakarshitha and Topaloglu, Rasit Onur and Ghosh, Swaroop},
  booktitle={2021 IEEE European Test Symposium (ETS)},
  pages={1--10},
  year={2021},
  organization={IEEE}
}

@article{benchmarking2023,
  title={Benchmarking Quantum Computers and the Impact of Quantum Noise},
  author={Salonik Resch and Ulya R. Karpuzcu},
  journal={ACM Computing Surveys},
  year={2021},
  doi={10.1145/3464420}
}

@phdthesis{disparatenoise2024,
  title={The Disparate Impact of Noise on Quantum Learning Algorithms},
  author={Armando Angrisani},
  school={Sorbonne Université},
  year={2023},
  note={Doctoral dissertation},
  url={https://theses.hal.science/tel-04511706v1}
}

@article{zneunified2023,
  title={Quantum Computing and Its Implications for Cybersecurity: A Comprehensive Review of Emerging Threats and Defenses},
  author={Sadik Khan and Krishnamoorthy Palani and Mrinal Goswami and Mohammed Salman Arafath},
  journal={Nanotechnology Perceptions},
  year={2024},
  doi={10.62441/nano-ntp.v20iS13.79}
}

@inproceedings{kashif2024hqnet,
  title={Hqnet: Harnessing quantum noise for effective training of quantum neural networks in nisq era},
  author={Kashif, Muhammad and Shafique, Muhammad},
  booktitle={2025 IEEE International Conference on Quantum Artificial Intelligence (QAI)},
  pages={387--394},
  year={2025},
  organization={IEEE}
}

@inproceedings{kashif2024nrqnn,
  title={NRQNN: The Role of Observable Selection in Noise-Resilient Quantum Neural Networks},
  author={Kashif, Muhammad and Shafique, Muhammad},
  booktitle={World Congress in Computer Science, Computer Engineering \& Applied Computing},
  pages={116--131},
  year={2024},
  organization={Springer}
}

@article{kashif2025deep,
  title={Deep quanvolutional neural networks with enhanced trainability and gradient propagation},
  author={Kashif, Muhammad and Shafique, Muhammad},
  journal={Scientific Reports},
  volume={15},
  number={1},
  pages={21764},
  year={2025},
  publisher={Nature Publishing Group UK London}
}

@article{biamonte2018qml,
  title={Quantum machine learning},
  author={Biamonte, Jacob and Wittek, Peter and Pancotti, Nicola and Rebentrost, Patrick and Wiebe, Nathan and Lloyd, Seth},
  journal={Nature},
  volume={549},
  number={7671},
  pages={195--202},
  year={2017},
  publisher={Nature Publishing Group UK London}
}

@article{abbas2020power,
  title={The power of quantum neural networks},
  author={Abbas, Amira and Sutter, David and Zoufal, Christa and Lucchi, Aur{\'e}lien and Figalli, Alessio and Woerner, Stefan},
  journal={Nature computational science},
  volume={1},
  number={6},
  pages={403--409},
  year={2021},
  publisher={Nature Publishing Group US New York}
}

@article{jeswal2018qnnreview,
  author    = {S. K. Jeswal and S. Chakraverty},
  title     = {Recent Developments and Applications in Quantum Neural Network: A Review},
  journal   = {Archives of Computational Methods in Engineering},
  year      = {2018},
  doi       = {10.1007/s11831-018-9269-0},
  publisher = {CIMNE, Barcelona, Spain}
}

@article{russo2024layerwise,
  title={Quantum error mitigation by layerwise Richardson extrapolation},
  author={Russo, Vincent and Mari, Andrea},
  journal={Physical Review A},
  volume={110},
  number={6},
  pages={062420},
  year={2024},
  publisher={APS}
}

@article{kashif2021design,
  title={Design space exploration of hybrid quantum--classical neural networks},
  author={Kashif, Muhammad and Al-Kuwari, Saif},
  journal={Electronics},
  volume={10},
  number={23},
  pages={2980},
  year={2021},
  publisher={MDPI}
}

@article{krebsbach2022optimized,
  title={Optimization of Richardson extrapolation for quantum error mitigation},
  author={Krebsbach, Michael and Trauzettel, Bj{\"o}rn and Calzona, Alessio},
  journal={Physical Review A},
  volume={106},
  number={6},
  pages={062436},
  year={2022},
  publisher={APS}
}

@article{vandenberg2022pec_sparse,
  title={Probabilistic error cancellation with sparse Pauli--Lindblad models on noisy quantum processors},
  author={Van Den Berg, Ewout and Minev, Zlatko K and Kandala, Abhinav and Temme, Kristan},
  journal={Nature physics},
  volume={19},
  number={8},
  pages={1116--1121},
  year={2023},
  publisher={Nature Publishing Group UK London}
}

@article{ezzell2022dd_survey,
  title={Dynamical decoupling for superconducting qubits: A performance survey},
  author={Ezzell, Nic and Pokharel, Bibek and Tewala, Lina and Quiroz, Gregory and Lidar, Daniel A},
  journal={Physical Review Applied},
  volume={20},
  number={6},
  pages={064027},
  year={2023},
  publisher={APS}
}

@article{cai2022quantumerrormitigation,
  title={Quantum error mitigation},
  author={Cai, Zhenyu and others},
    author2={Cai, Zhenyu and Babbush, Ryan and Benjamin, Simon C and Endo, Suguru and Huggins, William J and Li, Ying and McClean, Jarrod R and O’Brien, Thomas E},
  journal={Reviews of Modern Physics},
  volume={95},
  number={4},
  pages={045005},
  year={2023},
  publisher={APS}
}

@article{ddd2024,
  title={Synergistic dynamical decoupling and circuit design for enhanced algorithm performance on near-term quantum devices},
  author={Ji, Yanjun and Polian, Ilia},
  journal={Entropy},
  volume={26},
  number={7},
  pages={586},
  year={2024},
  publisher={MDPI}
}

@misc{mitiqdocs,
  author       = {Unitary Fund and Contributors},
  title        = {Mitiq Documentation: A Python toolkit for quantum error mitigation},
  howpublished = {\url{https://mitiq.readthedocs.io/en/stable/}},
  note         = {Accessed: 2025-07-03},
  year         = {2025}
}

@inproceedings{giurgica2021zne,
  title={Digital zero noise extrapolation for quantum error mitigation},
  author={Giurgica-Tiron, Tudor and Hindy, Yousef and LaRose, Ryan and Mari, Andrea and Zeng, William J},
  booktitle={2020 IEEE international conference on quantum computing and engineering (QCE)},
  pages={306--316},
  year={2020},
  organization={IEEE}
}

@article{zhang2023benchmarking,
  title={Benchmarking noisy intermediate scale quantum error mitigation strategies for ground state preparation of the HCl molecule},
  author={Weaving, Tim and Ralli, Alexis and Kirby, William M and Love, Peter J and Succi, Sauro and Coveney, Peter V},
  journal={Physical Review Research},
  volume={5},
  number={4},
  pages={043054},
  year={2023},
  publisher={APS}
}

@inproceedings{ahmed2025noisyhqnn,
  title={Noisy hqnns: A comprehensive analysis of noise robustness in hybrid quantum neural networks},
  author={Ahmed, Tasnim and Marchisio, Alberto and Kashif, Muhammad and Shafique, Muhammad},
  booktitle={2025 International Joint Conference on Neural Networks (IJCNN)},
  pages={1--10},
  year={2025},
  organization={IEEE}
}

@article{kashif2024resqnets,
  title={ResQNets: a residual approach for mitigating barren plateaus in quantum neural networks},
  author={Kashif, Muhammad and Al-Kuwari, Saif},
  journal={EPJ Quantum Technology},
  volume={11},
  number={1},
  pages={1--28},
  year={2024},
  publisher={SpringerOpen}
}

@misc{Dua:2019,
  author = {Dua, Dheeru and Graff, Casey},
  title = {{UCI} Machine Learning Repository},
  year = {2019},
  url = {http://archive.ics.uci.edu/ml},
  DOI = {10.24432/C56C76},
  institution = {University of California, Irvine, School of Information and Computer Sciences}
}

@article{innan2025optimizing,
  title={Optimizing low-energy carbon iiot systems with quantum algorithms: Performance evaluation and noise robustness},
  author={Dave, Kshitij and Innan, Nouhaila and Behera, Bikash K and Mumtaz, Shahid and Al-Kuwari, Saif and Farouk, Ahmed},
  journal={IEEE Internet of Things Journal},
  volume={12},
  number={17},
  pages={34653--34662},
  year={2025},
  publisher={IEEE}
}

@article{innan2024financial,
  title={Financial fraud detection using quantum graph neural networks},
  author1={Innan, Nouhaila and others},
    author={Innan, Nouhaila and Sawaika, Abhishek and Dhor, Ashim and Dutta, Siddhant and Thota, Sairupa and Gokal, Husayn and Patel, Nandan and Khan, Muhammad Al-Zafar and Theodonis, Ioannis and Bennai, Mohamed},
  journal={Quantum Machine Intelligence},
  volume={6},
  number={1},
  pages={7},
  year={2024},
  publisher={Springer}
}

@article{innan2025qnn,
  title={Qnn-vrcs: A quantum neural network for vehicle road cooperation systems},
  author={Innan, Nouhaila and Behera, Bikash K and Al-Kuwari, Saif and Farouk, Ahmed},
  journal={IEEE Transactions on Intelligent Transportation Systems},
  year={2025},
  publisher={IEEE}
}

@inproceedings{innan2025lep,
  title={Lep-qnn: Loan eligibility prediction using quantum neural networks},
  author={Innan, Nouhaila and Marchisio, Alberto and Bennai, Mohamed and Shafique, Muhammad},
  booktitle={2025 IEEE International Conference on Quantum Computing and Engineering (QCE)},
  volume={1},
  pages={1864--1872},
  year={2025},
  organization={IEEE}
}

@article{innan2024quantum,
  title={Quantum state tomography using quantum machine learning},
  author1={Innan, Nouhaila and others},
    author={Innan, Nouhaila and Siddiqui, Owais Ishtiaq and Arora, Shivang and Ghosh, Tamojit and Ko{\c{c}}ak, Yasemin Poyraz and Paragas, Dominic and Galib, Abdullah Al Omar and Khan, Muhammad Al-Zafar and Bennai, Mohamed},
  journal={Quantum Machine Intelligence},
  volume={6},
  number={1},
  pages={28},
  year={2024},
  publisher={Springer}
}

@inproceedings{innan2025next,
  title={Next-Generation Quantum Neural Networks: Enhancing Efficiency, Security, and Privacy},
  author={Innan, Nouhaila and Kashif, Muhammad and Marchisio, Alberto and Bennai, Mohamed and Shafique, Muhammad},
  booktitle={2025 IEEE 31st International Symposium on On-Line Testing and Robust System Design (IOLTS)},
  pages={1--4},
  year={2025},
  organization={IEEE}
}

@article{el2026comparative,
  title={Comparative performance analysis of quantum machine learning architectures for credit card fraud detection},
  author={El Alami, Mansour and Innan, Nouhaila and Shafique, Muhammad and Bennai, Mohamed},
  journal={Applied Intelligence},
  volume={56},
  number={3},
  pages={83},
  year={2026},
  publisher={Springer}
}

@article{dave2025sentiqnf,
  title={Sentiqnf: A novel approach to sentiment analysis using quantum algorithms and neuro-fuzzy systems},
  author={Dave, Kshitij and Innan, Nouhaila and Behera, Bikash K and Mumtaz, Zahid and Al-Kuwari, Saif and Farouk, Ahmed},
  journal={IEEE Transactions on Computational Social Systems},
  year={2025},
  publisher={IEEE}
}

@inproceedings{pathak2024resource,
  title={Resource Allocation Optimization in 5G Networks Using Variational Quantum Regressor},
  author={Pathak, Param and Oad, Vidhi and Prajapati, Aditya and Innan, Nouhaila},
  booktitle={2024 International Conference on Quantum Communications, Networking, and Computing (QCNC)},
  pages={101--105},
  year={2024},
  organization={IEEE}
}

@article{dutta2025quiet,
  title={QUIET-SR: Quantum Image Enhancement Transformer for Single Image Super-Resolution},
  author={Dutta, Siddhant and Innan, Nouhaila and Najafi, Khadijeh and Yahia, Sadok Ben and Shafique, Muhammad},
  journal={arXiv preprint arXiv:2503.08759},
  year={2025}
}

@INPROCEEDINGS{11228312,
  author={Siddiqui, Owais Ishtiaq and Innan, Nouhaila and Marchisio, Alberto and Bennai, Mohamed and Shafique, Muhammad},
  booktitle={2025 International Joint Conference on Neural Networks (IJCNN)}, 
  title={Quantum Bayesian Networks for Machine Learning in Oil-Spill Detection}, 
  year={2025},
  volume={},
  number={},
  pages={1-8},
  doi1={10.1109/IJCNN64981.2025.11228312}}

@article{choudhary2025hqnn,
  title={{HQNN-FSP}: A Hybrid Classical-Quantum Neural Network for Regression-Based Financial Stock Market Prediction},
  author={Choudhary, Prashant Kumar and Innan, Nouhaila and Shafique, Muhammad and Singh, Rajeev},
  journal={arXiv preprint arXiv:2503.15403},
  year={2025}
}

@article{atallah2025investigating,
  title={Investigating Different Barren Plateaus Mitigation Strategies in Variational Quantum Eigensolver},
  author={Atallah, Mostafa and Innan, Nouhaila and Kashif, Muhammad and Shafique, Muhammad},
  journal={arXiv preprint arXiv:2512.11171},
  year={2025}
}

\end{document}